\documentclass{webofc}
\usepackage[frozencache=true, cachedir=minted-cache]{minted}
\usepackage{wasysym}
\usepackage[varg]{txfonts}   

\newcommand\run{\textit{Run}}
\newcommand\subrun{\textit{Subrun}}
\newcommand\event{\textit{Event}}
\newcommand\fold[1]{\,{}^#1\!/\,}
\newcommand\unfold[1]{\,{}_#1\!\backslash\,}

\newcommand\art{\textit{art}\xspace}
\newcommand\makeoffset{\texttt{make\_offset}\xspace}
\newcommand\maketracks{\texttt{make\_tracks}\xspace}
\newcommand\goodhits{\texttt{"GoodHits"}\xspace}
\newcommand\goodtracks{\texttt{"GoodTracks"}\xspace}
\newcommand\calibrationentry{\texttt{"CalibrationEntry"}\xspace}
\newcommand\calibrationoffset{\texttt{"CalibrationOffset"}\xspace}

\begin{document}

\definecolor{bg}{rgb}{0.95,0.95,0.95}

\title{Meld}
\subtitle{Exploring the Feasibility of a Framework-less Framework}

\author{\firstname{Kyle} \lastname{Knoepfel}\inst{1}\fnsep\thanks{\email{knoepfel@fnal.gov}}}

\institute{Fermi National Accelerator Laboratory, P.O. Box 500, Batavia, Illinois, United States}

\abstract{
HEP data-processing frameworks are essential ingredients in getting from raw data to physics results. But they are often tricky to use well, and they present a significant learning barrier for the beginning HEP physicist. In addition, existing frameworks typically support rigid, collider-based data models, which do not map well to neutrino-physics experiments like DUNE. Neutrino physicists thus expend significant effort working around framework limitations instead of using a framework that directly supports their needs. \medskip

Presented here is Meld, a Fermilab R\&D project, which intends to address these limitations. By leveraging modern C++ capabilities, state-of-the-art concurrency libraries, and a flexible data model, it is possible for beginning (and seasoned) HEP physicists to execute framework programs easily and efficiently, with minimal coupling to framework-specific constructs. Meld aims to directly support the frameworks needs of neutrino experiments like DUNE as well as the more common collider-based experiments.
}

\maketitle

\section{Introduction}
\label{sec:intro}
The Deep Underground Neutrino Experiment (DUNE) is Fermilab's flagship scientific effort toward further understanding neutrino oscillations, measuring potential charge-parity violation in the neutrino sector, and detecting neutrinos from supernova bursts and beyond-the-Standard Model physics.  The experiment's near detector is located at Fermilab, and the far detector is composed of one or more liquid Argon (LAr) time-projection chambers (TPCs) at the Sanford Underground Research Facility near Lead, South Dakota.  The LAr active volume in the TPCs affords very high resolution in the detection of neutrino interactions.

Like other experiments, DUNE will use an offline computing framework to process the data collected by the online data acquisition system.  However, whereas LHC experiments rely on hadronic collisions occurring within localized accelerator bunch crossings, the neutrino interactions recorded by DUNE will be distributed throughout very large detector volumes.  Unfortunately, most reconstruction and simulation frameworks in HEP (e.g. see~\cite{bib:atlas,bib:cms,bib:o2,bib:art}) assume collider-physics concepts that are not always appropriate for neutrino experiments.  Even DUNE's current offline framework \art~\cite{bib:art} reflects a collider-centric design as it originated as a fork of the CMS framework.  A framework is therefore needed that can support DUNE without physicists having to work around the collider assumptions imposed by existing frameworks.

The DUNE experiment has formulated a list of offline framework requirements~\cite{bib:offline-cdr} based on the physics goals mentioned above.  Some of those requirements state that:
\begin{itemize}
    \item physics algorithms should be framework-agnostic,
    \item the framework must be able to break apart events into smaller chunks for more granular processing, and then stitch those chunks back together into an event,
    \item the framework should support ``sliding event windows'' to provide ``edge effect'' coverage for extended time readouts during supernovae events, and
    \item the framework should make minimal assumptions about the data model.
\end{itemize}
To determine the feasibility of developing a framework that could meet such requirements, a laboratory-directed R\&D project, now called Meld, was established in 2022 at Fermilab.

This article explains Meld's design by first discussing in Section~\ref{sec:logical} the logical foundations of data products, data hierarchies, and how data products are processed within those hierarchies.  Section~\ref{sec:higher-order-functions} presents the concept of higher-order functions and their roles within a dependency graph of user-defined operations.  The consequences of a functional approach on interfaces are addressed in Section~\ref{sec:declarative}, which is followed in Section~\ref{sec:implementation} by a description of the features supported by Meld's prototype implementation.

\section{Logical representations of data}
\label{sec:logical}

Reconstruction and simulation frameworks in HEP refer to their managed data objects as \textit{data products}, each of which is:
\begin{enumerate}
    \item Identifiable, to facilitate user-specification of what data to analyze,
    \item A member of at least one data set (or \textit{domain}) such as an \textit{event},
    \item Immutable, as implied by the definition of an element of a mathematical set, and
    \item Opaque to the framework, implying a separation of the user space from the framework.
\end{enumerate}

Data products can be created from other data products through user-defined functions or \textit{mappings}, the details of which are defined based on the use case.  In the case of \art, users do not create explicit mappings between data products but instead apply implicit mappings while interacting directly with the domain containing the data product.  One of the goals of Meld is to replace implicit mappings between data products with explicit mappings that do not directly depend on the data-product domain.

\subsection{Data hierarchies}
\label{sec:data-hierarchies}

As mentioned above, data products do not exist by themselves, but they are members of a domain.  Such domains are often called events, but as events are often contained by supersets (such as \textit{luminosity sections}, \textit{subruns}, or \textit{runs}), then a given event data product logically is also a member of any supersets of that event.  The organization of data sets and their relationships with each other define a \textit{data hierarchy}.

Commonly used data hierarchies include tree-like structures that honor relations such as $\run_i \supset \subrun_j \supset \event_k$, which allows an unambiguous identification of an event through the number triplet $(i, j, k)$.  Although frameworks may allow an experiment to specify the range of values taken by the identifiers $i$,  $j$, and $k$, the hierarchical relation is rigid, thus constraining programs to data organizations that may not match well the processing needs of the user.  The DUNE experiment has found this to be the case, where the lack of granularity below the event domain yields processing complications and program-memory footprints that are difficult to manage.  To satisfy DUNE's needs, a system is required that can support dynamic and more general hierarchies than those discussed above (some examples are shown in Section~\ref{sec:implementation}).

\subsection{Processing data in two ways}
\label{sec:functional-processing}

\begin{figure*}
\begin{minipage}[c][6.5cm]{0.618\textwidth}
    \centering
    \includegraphics[height=6cm]{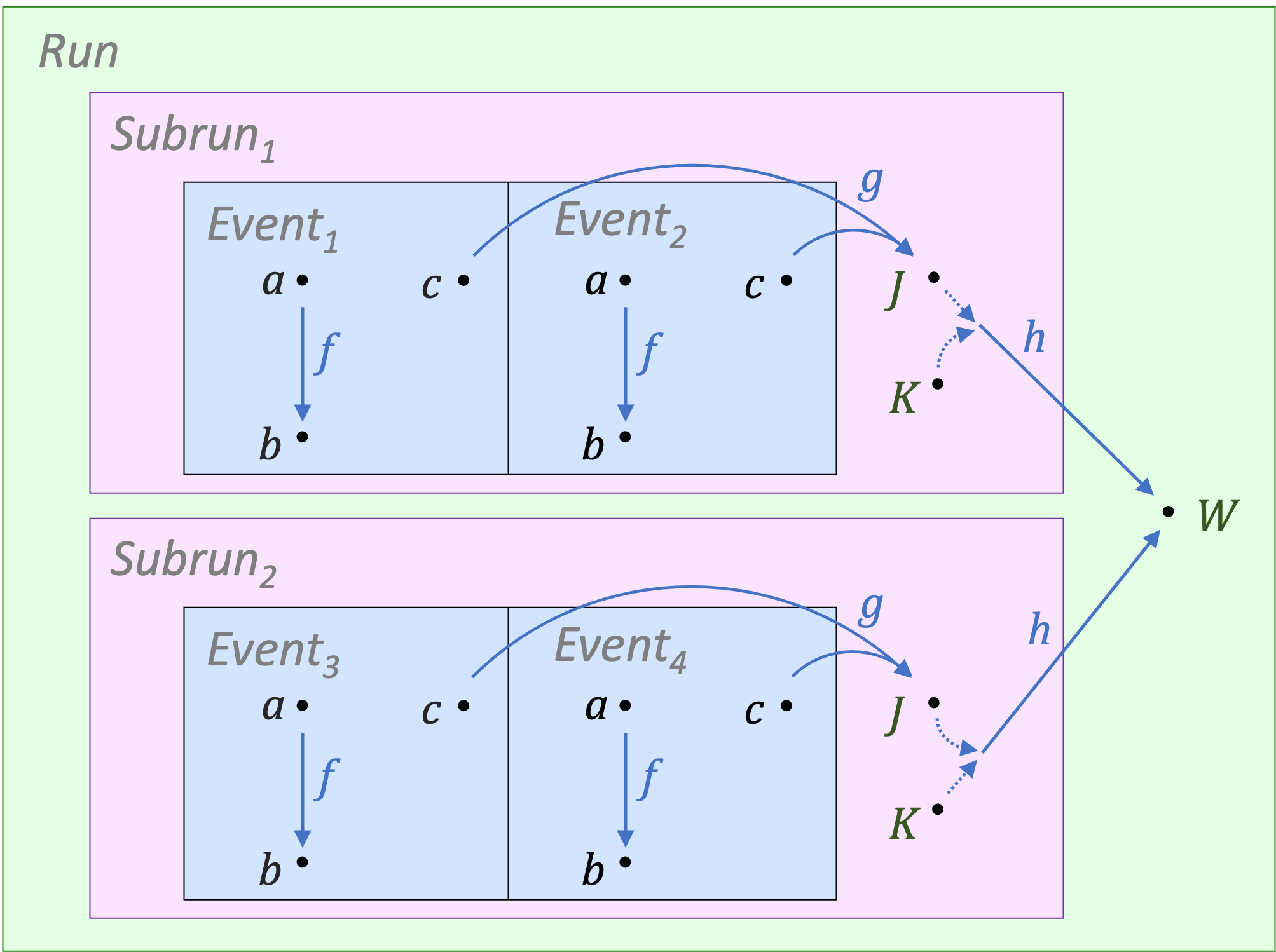}
    \small (a) Procedural approach
\end{minipage} \quad %
\begin{minipage}[c][6.5cm]{0.35\textwidth}
    \centering
    \includegraphics[height=6cm]{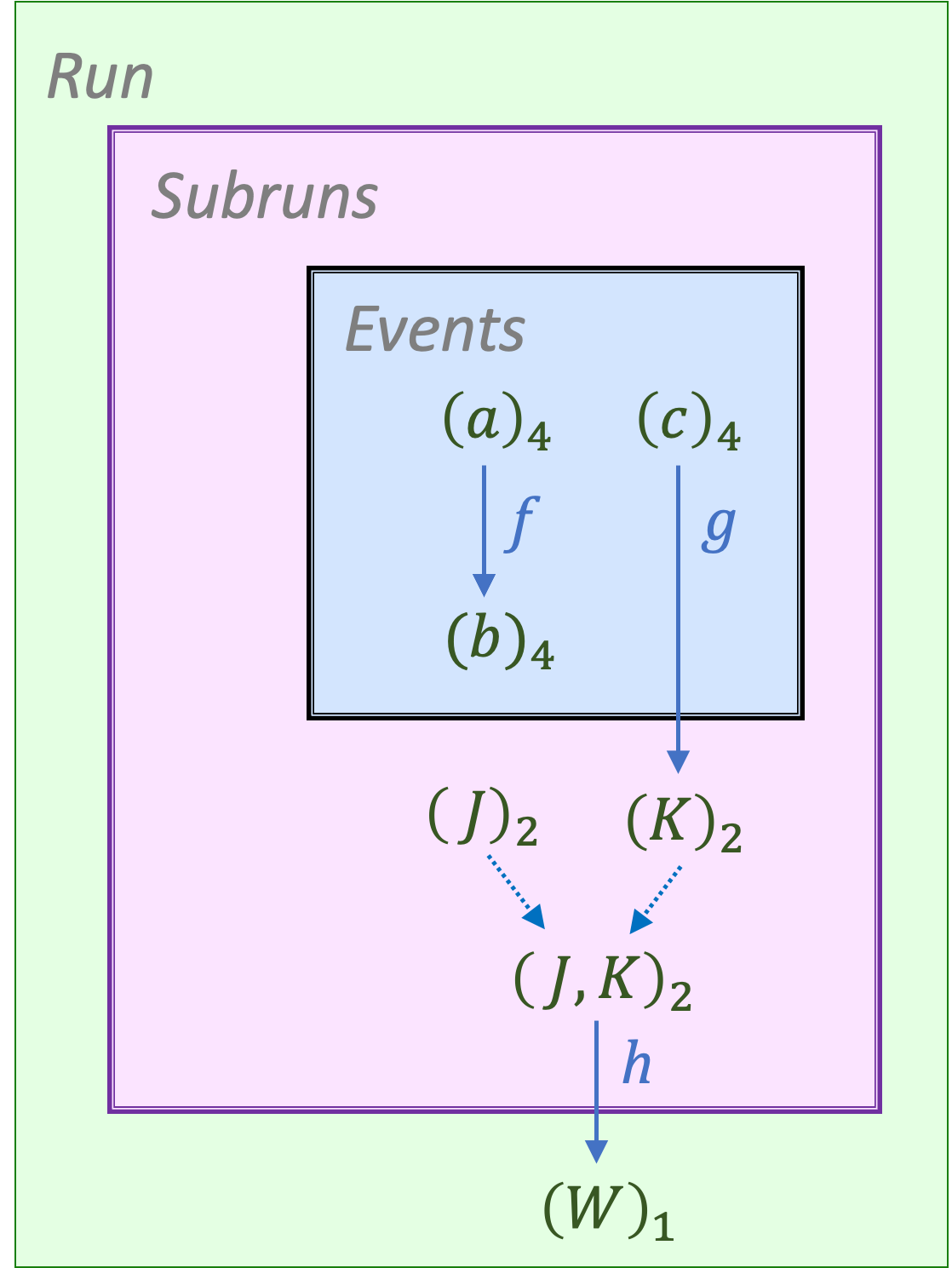}
    \small(b) Functional approach
\end{minipage}
\caption{Logical representations of processing data with (a) a procedural approach, and (b) a functional approach.}
\label{fig:data-processing}
\end{figure*}

Figure~\ref{fig:data-processing} depicts two ways of processing a tree-like data set that contains one run, two subruns within the run, and two events within each subrun (four events total).  The run contains a data product (i.e. an object) labeled $W$, each subrun contains two data products $J$ and $K$, and each event contains the data products $a$, $b$, and $c$.  The mappings between the data products include:
\begin{itemize}
    \item $f: a \mapsto b$ for each event,
    \item $g: c \mapsto J$ for each event such that only one data product $J$ is created for each subrun, and
    \item $h: (J, K) \mapsto W$ for each subrun such that only one run data product $W$ is created.
\end{itemize}

Whereas $f$ is a simple mapping or \textit{transform} of one data product to another within an event, $g$ and $h$ correspond to \textit{reductions} (or \textit{folds}) where multiple data products across multiple domains are aggregated into one data product of an encapsulating domain.  Examples of reductions within the HEP community include histograms and event counts, which are commonplace ingredients used to obtain physics results.

Panel (a) of Figure~\ref{fig:data-processing} shows a procedural way of processing data, where the domain of each data product is explicit as is each invocation of the user-defined mappings.  This approach is common among HEP frameworks, which often require users to interact with the domain containing the data instead of transparently providing the requested data.  A consequence of this programming paradigm is that users tailor their code to work with stateful objects that represent the domains.  The significant boilerplate code required for such a pattern often leads to poor separation of physics algorithms (which generally can be framework-agnostic) from the code required to register those algorithms with the framework.

Meld, however, adopts a functional processing paradigm, as illustrated in panel (b) of Figure~\ref{fig:data-processing}.  With a functional approach, the framework forms sequences of data products across common domains (e.g. all events) and applies the user-defined mapping to each data product of the sequence.  For example, the mapping $f$ is applied to each data product $a$ contained in events 1 through 4.  Using a notation similar to Ref.~\cite{bib:bird}, this operation can be expressed using a \textit{higher-order function}:
\begin{equation}\label{eqn:f}
    f \ast (a)_4 = (b)_4
\end{equation}
where the $\ast$ infix operator accepts a function $f$ and the sequence $(a)_4$, which signifies a sequence of 4 data products, each labeled $a$.  The framework still records the domain of each data product, but the domain is retained by the framework as a bookkeeping artifact that is accessible by the user only when necessary.

Note that the formation of data-product sequences in the functional approach need not be ``strict'' in the sense that all data products of a sequence must exist before the framework can apply the user-defined mapping.  In some cases (e.g. a data acquisition environment), the full sequence cannot be known at any given time, and even when the full sequence can be known, memory limitations of the machine may constrain the framework to process only a subset of the sequence at a time.

\section{Higher-order functions and graphs}
\label{sec:higher-order-functions}

As mentioned above, Equation~\ref{eqn:f} gives an example of a higher-order function, which is any function that (a) receives at least one function as one of its function parameters, or (b) returns a function.  Case (a) applies in framework contexts, where the user provides a callable entity or \textit{operator} to be applied to sequence elements, specifies the pattern used to form that data-product sequence, and chooses the kind of higher-order function desired (e.g. transform).

Table~\ref{tab:higher-order-functions} lists the higher-order functions for some computing patterns commonly used within HEP and the broader computing industry.  Although it is typical among HEP applications to transform, filter, monitor, and reduce data products or their domains, it is less common to split them.  However, as mentioned in Section~\ref{sec:intro}, the splitting of data-product domains is one of DUNE's framework requirements.  Meld was therefore designed to support such splitting.

Framework jobs generally consist of many user-defined functions that must be applied to data in a particular composition order.  This order results in a dependency graph that can be exploited to optimize the processing of the data products.  Both panels in Figure~\ref{fig:data-processing} can be considered dependency graphs.  However, whereas panel (a) depicts a graph of many identically named nodes and edges, panel (b) shows a significantly smaller graph where nodes and edges have unique names.  A functional approach thus enables a simpler graphical description of the processing required than the procedural approach.

\begin{table*}
\centering
\caption{Higher-order functions and their operator signatures.  For each pattern, $a \in \alpha$ and $b \in \beta$, where $\alpha$ and $\beta$ are mathematical sets representing data products.  The mathematical set $\mathbf{1}$ (or \texttt{void} in the C and C++ languages) contains one element called \textit{void}.  The return type of $\mathbf{1} + \alpha \times \beta$ denotes a \textit{tagged union} whereby the return value can either be \textit{void} or of the set $\alpha \times \beta$.}
\label{tab:higher-order-functions}
\begin{tabular}{lll}
\hline
Computing pattern & Higher-order function & Operator signature  \\\hline
Transform (map) & $f \ast (a)_n = (b)_n$ & $f: \alpha \rightarrow \beta$ \\
Filter & $f \lhd (a)_n = (a)_m$ where $m \le n$ & $f: \alpha \rightarrow \mathrm{Boolean}$ \\
Monitor (absorb) & $f \LHD (a)_n = (\ )$ & $f: \alpha \rightarrow \mathbf{1}$\\
Reduce (fold) & $f \fold{n} (a)_n = b$ & $f: \beta \times \alpha \rightarrow \beta$ \\
Split (unfold) & $f \unfold{n} b = (a)_n$ & $f: \beta \rightarrow \mathbf{1} + \alpha \times \beta$ \\
\hline
\end{tabular}
\end{table*}

\subsection{Side effects and concurrency support}

The ability to support concurrent processing of data depends strongly on how many side effects are present in the program (e.g. accessing global state,  or manipulating objects shared among threads).  Table~\ref{tab:higher-order-functions} assumes that the user-defined operators contain no side effects, thus making the invocation of such operators automatically safe within a multi-threaded or multi-process context.

The functional approach does permit some stateful objects, such as the result object of a reduction (e.g. a histogram).  However, it is possible in some cases for the framework to manage and update the result object in a safe manner without placing that burden on the framework user.

\section{Declarative interfaces}
\label{sec:declarative}

By treating framework jobs as mappings of data products through higher-order functions, a declarative user interface is possible.  Figure~\ref{fig:listing1} shows such an interface where a user-defined, framework-agnostic physics algorithm called \maketracks is registered with Meld through the \texttt{DEFINE\_MODULE} C++ preprocessor invocation.  The registration occurs at runtime (upon loading the corresponding shared object library), instructing the framework to apply the user-defined \maketracks function to the \goodhits data product in each event and to then store the result as the \goodtracks data product.

The framework uses type deduction to match the input and output data product names with their corresponding C++ types---namely, the \goodhits data product is presented as type \texttt{Hits} to the user, and the return value of type \texttt{Tracks} is stored as the \goodtracks data product.  The user may specify the allowed concurrency level when invoking the function---an unsigned integer that represents the number of data products that can be processed concurrently by the framework's execution of the registered function.  The special keyword \texttt{unlimited} instructs the framework that it may use the maximum available concurrency provided by the system on which the \maketracks function is invoked.

Understandably, the user may not wish to hard-code the data-product labels (\goodhits and \goodtracks), the domain name (\texttt{"Event"}), and the allowed concurrency number.  In such cases, the \texttt{config} object may be queried for the value of a configuration parameter specified by the user at runtime.

\subsection{More complicated models}
\label{sec:realistic}

An objection can be made that the example shown in Figure~\ref{fig:listing1} is too simplistic to be useful in a realistic HEP computing context.  An important capability of Meld is, therefore, the ability to register with the framework functions that receive arguments corresponding to data products from different domains.  Figure~\ref{fig:listing2} shows such an example, where the \maketracks function receives an additional function parameter with calibration information provided through the \calibrationentry run data product.

In some cases, the \maketracks function may require only an object (e.g. an offset) that is created from a \calibrationentry data product and not the full calibration information.  If creating that offset is time-consuming, it is undesirable to repeatedly create the object for each invocation of \maketracks.  The user can therefore register a separate \makeoffset function with the framework, which creates the offset once for each run.  That offset is then available as a data product that can be used by the \maketracks function.  Based on the data products required and produced by each algorithm, the framework ensures that algorithms are processed in the correct order.  Figure~\ref{fig:listing3} shows this use case, using the \texttt{to\_temporary} clause to denote a data product that should not be persisted to disk.

\begin{figure}
\small
\begin{minted}[bgcolor=bg]{C++}
#include "meld/module.hpp"

namespace expt {
  Tracks make_tracks(Hits const& hits) { ... }
}

DEFINE_MODULE(m, config) {
  m.with(expt::make_tracks, concurrency::unlimited)
   .transform("GoodHits").in_each("Event").to("GoodTracks");
}
\end{minted}
\caption{Example of registering a user-defined function \maketracks
with Meld, such that all event data products labeled \goodhits are transformed to data products labeled \goodtracks.  More complicated scenarios are described in Section~\ref{sec:realistic}.}
\label{fig:listing1}

\begin{minted}[bgcolor=bg]{C++}
#include "meld/module.hpp"

namespace expt {
  Tracks make_tracks(Hits const& hits, Entry const& entry) { ... }
}

DEFINE_MODULE(m, config) {
  m.with(expt::make_tracks, concurrency::unlimited)
   .transform("GoodHits"_in_each("Event"), "CalibrationEntry"_in_each("Run"))
   .to("GoodTracks");
}
\end{minted}
\caption{Similar scenario as Figure~\ref{fig:listing1}, but using the \calibrationentry run data product as an additional function argument to \maketracks.}
\label{fig:listing2}

\begin{minted}[bgcolor=bg]{C++}
#include "meld/module.hpp"

namespace expt {
  Offset make_offset(Entry const& entry) { ... }
  Tracks make_tracks(Hits const& hits, Offset const& offset) { ... }
}

DEFINE_MODULE(m, config) {
  m.with(expt::make_offset, concurrency::unlimited)
   .transform("CalibrationEntry").in_each("Run")
   .to_temporary("CalibrationOffset");
  m.with(expt::make_tracks, concurrency::unlimited)
   .transform("GoodHits"_in_each("Event"), "CalibrationOffset"_in_each("Run"))
   .to("GoodTracks");
}
\end{minted}
\caption{Similar scenario as Figure~\ref{fig:listing2}, but a separate function \makeoffset is used to perform the calculation of \calibrationoffset once per run instead of once per event.}
\label{fig:listing3}
\end{figure}

\section{Meld implementation and execution backend}
\label{sec:implementation}

Meld itself does not prescribe a particular implementation, but a C++ prototype is available~\cite{bib:meld} that uses Intel's oneTBB flow graph library~\cite{bib:oneTBB} as the multi-threading backend.  Each of the computing patterns listed in Table~\ref{tab:higher-order-functions} is supported by the prototype.  In addition, a zip utility is provided so that multiple data products can be presented to a user-defined function that takes more than one argument (e.g. the $h$ reduction in Section~\ref{sec:functional-processing}).

\begin{figure*}
\label{fig:data-hierarchies}
\begin{minipage}{0.31\textwidth}
    \centering
    \includegraphics[width=\textwidth]{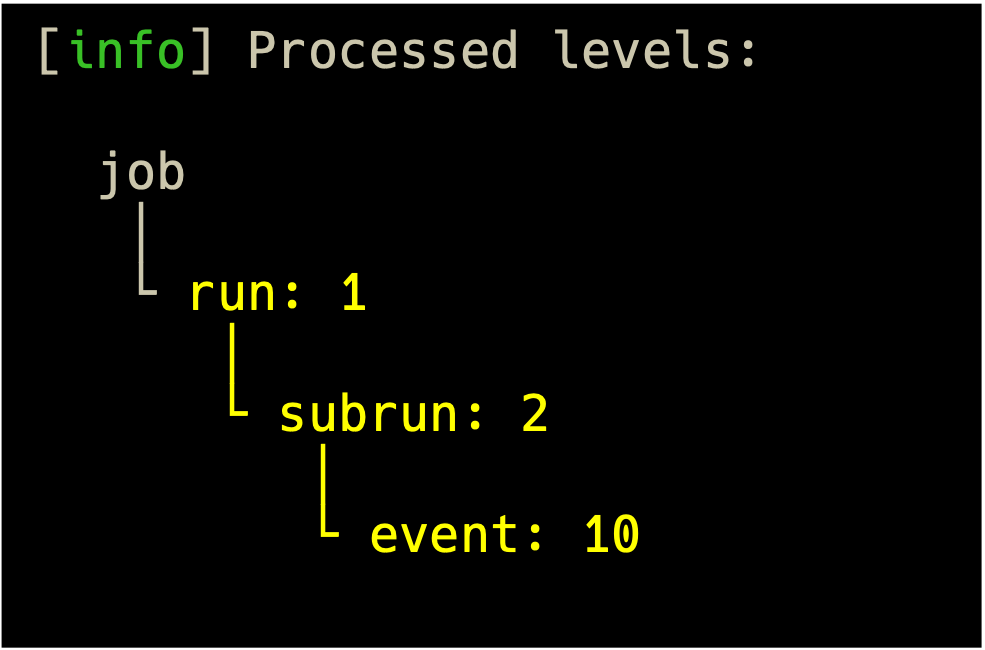}
    \small (a) \art-based hierarchy
\end{minipage} \quad %
\begin{minipage}{0.31\textwidth}
    \centering
    \includegraphics[width=\textwidth]{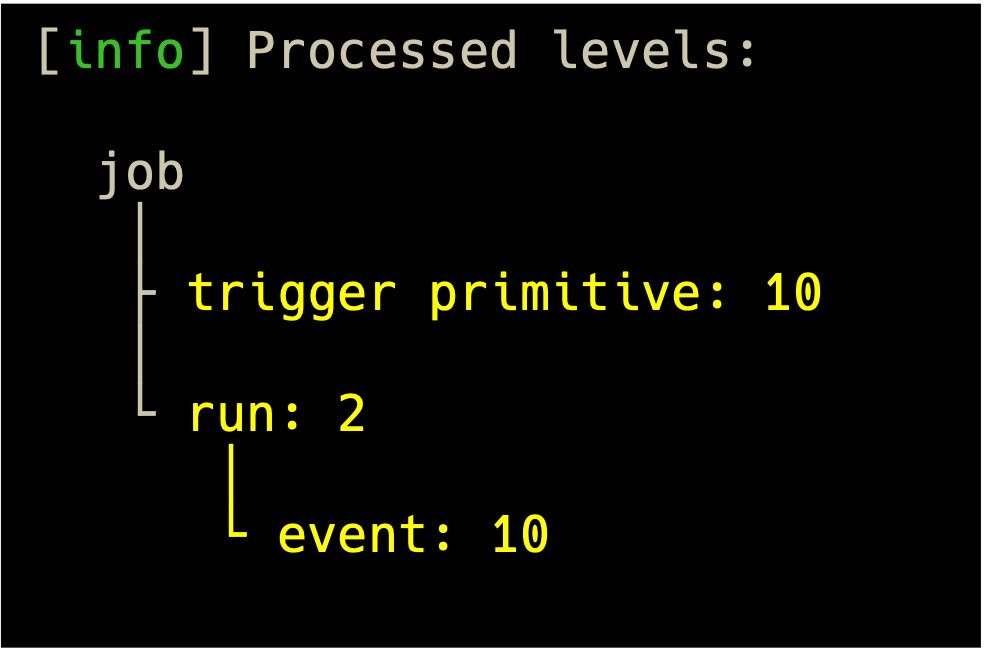}
    \small (b) Non-trivial hierarchy
\end{minipage} \quad %
\begin{minipage}{0.31\textwidth}
    \centering
    \vspace{1cm}
    \includegraphics[width=\textwidth]{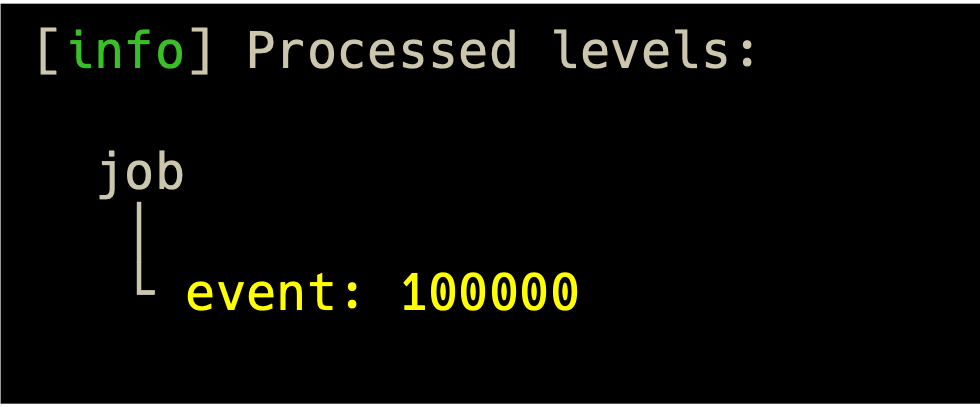}
    \small (c) Flat hierarchy
\end{minipage}
\caption{Sample hierarchies supported by Meld.  The numbers for nested levels correspond to aggregations across all supersets (e.g. for the \art-based hierarchy, a total of 10 events across 2 subruns were processed).}
\end{figure*}

Figure~\ref{fig:data-hierarchies} shows three sample hierarchies that Meld supports.  The hierarchy summaries are printed to the terminal at the end of each Meld program, which records the domains processed and their relations to each other.  Hierarchy (a) follows the tree-like structure (see Section~\ref{sec:data-hierarchies}) currently used by the \art framework.  Hierarchy (b) is non-trivial in that runs and trigger primitives are orthogonal domains, but events are subsets of runs.  The flat hierarchy (c) processes only events, which are contained only by the ``job'' domain.

Features yet to be supported by Meld include:
\begin{itemize}
    \item a ``sliding window" over domains, where multiple data products from different domains can be presented to the user at the same time,
    \item a conditions system where calibration constants do not necessarily belong to only one data hierarchy,
    \item asynchronous processing of algorithms executed on non-CPU resources (e.g. GPUs), and
    \item optimizations in scheduling the processing of time-ordered data \textit{vs.} that of a complete data set, whose elements are known ahead of time.
\end{itemize}

\section{Conclusion}

Meld's functional approach of processing data products enables a decomposition of a framework job into a granular set of basic computing idioms.  The presented declarative interface avoids significant entanglements between user code and framework boilerplate, resulting in a much simpler (even if not framework-less) framework model for experiments.  This avenue of exploration has been received favorably by members of the DUNE collaboration, and the successes achieved thus far suggest that higher-order functions have more of a place in HEP data-processing frameworks than has been previously recognized.

\section{Acknowledgments}

The author thanks the DUNE experiment for many fruitful discussions about their framework needs.  He is also grateful to Marc Paterno for his helpful input and careful reading of this manuscript, and for his insight into how functional programming paradigms are relevant to the HEP community.

This manuscript has been authored by Fermi Research Alliance, LLC under Contract No. DE-AC02-07CH11359 with the U.S. Department of Energy, Office of Science, Office of High Energy Physics.


\begin{thebibliography}{}

\bibitem{bib:atlas}
Charles Leggett, \textit{et al}, J. Phys. Conf. Ser. \textbf{898}, 042009 (2017)

\bibitem{bib:cms}
E. Sexton-Kennedy, \textit{et al}, J. Phys. Conf. Ser. \textbf{608}, 012034 (2015)   

\bibitem{bib:o2}
J. Adam, \textit{et al} [ALICE Collaboration], ``Technical Design Report for the Upgrade of the Online-Offline Computing System'', CERN-LHCC-2015-006, ALICE-TDR-019 (2015)

\bibitem{bib:art}
C. Green, \textit{et al}, J. Phys. Conf. Ser. \textbf{396}, 022020 (2012) 

\bibitem{bib:offline-cdr}
A. Abed Abud, \textit{et al} [DUNE collaboration], DUNE Offline Computing Conceptual Design Report, arXiv:2210.15665 [physics.data-an] (2022)

\bibitem{bib:bird}
R. Bird, Logic of Programming and Calculi of Discrete Design, NATO ASI Series \textbf{36} \url{https://doi.org/10.1007/978-3-642-87374-4_1} (1987) 

\bibitem{bib:meld}
Meld, GitHub repository, \url{https://github.com/knoepfel/meld} (2023)

\bibitem{bib:oneTBB}
oneAPI Threading Building Blocks (oneTBB), \url{https://oneapi-src.github.io/oneTBB}

\end{thebibliography}
\end{document}